\documentclass[12pt,preprint]{aastex}

\usepackage{amsmath}  
\usepackage{epsfig}  

\begin{document}
\bibliographystyle{apj}

\shortauthors{Chiang \& Choi}
\shorttitle{Kuiper Belt Plane}

\title{The Warped Plane of the Classical Kuiper Belt}
\author{Eugene~Chiang\altaffilmark{1} and Hyomin Choi\altaffilmark{2}}
\altaffiltext{1}{Center for Integrative Planetary Sciences,
Department of Astronomy,
University of California at Berkeley,
Berkeley, CA~94720, USA}
\altaffiltext{2}{Department of Mathematics,
University of California at Berkeley,
Berkeley, CA 94720, USA}
\email{echiang@astro.berkeley.edu, hyomin@berkeley.edu}


\begin{abstract}
  By numerically integrating the orbits of the giant planets and of
  test particles over a period of four billion years, we follow the evolution of
  the location of the midplane of the Kuiper belt. The Classical
  Kuiper belt conforms to a warped sheet that precesses with a 1.9 Myr
  period.  The present-day location of the Kuiper belt plane can be
  computed using linear secular perturbation theory: the local normal
  to the plane is given by the theory's forced inclination vector,
  which is specific to every semi-major axis. The Kuiper belt plane
  does not coincide with the invariable plane, but deviates from it by
  up to a few degrees in stable zones. For example, at a semimajor
  axis of 38 AU, the local Kuiper belt plane has an inclination of 1.9
  degrees and a longitude of ascending node of 149.9 degrees when
  referred to the mean ecliptic and equinox of J2000. At a semimajor
  axis of 43 AU, the local plane has an inclination of 1.9 degrees and
  a nodal longitude of 78.3 degrees. Only at infinite semimajor axis
  does the Kuiper belt plane merge with the invariable plane, whose
  inclination is 1.6 degrees and nodal longitude is 107.7 degrees. A
  Classical Kuiper belt object keeps its inclination relative to the
  Kuiper belt plane nearly constant, even while the latter plane departs from
  the trajectory predicted by linear theory. The constancy of relative
  inclination reflects the undamped amplitude of free oscillation;
  that is, the homogeneous solution to the forced harmonic oscillator
  equation retains constant amplitude, even while the inhomogeneous
  solution cannot be written down accurately because the planetary
  forcing terms are chaotic.  Current observations of Classical Kuiper
  belt objects are consistent with the plane being warped by the giant
  planets alone, but the sample size will need to increase by a few
  times before confirmation exceeds $3\sigma$ in confidence.  In
  principle, differences between the theoretically expected plane and
  the observed plane could be used to infer as yet unseen masses
  orbiting the Sun, but carrying out such a program would be
  challenging.
\end{abstract}

\section{INTRODUCTION}
\label{sec_intro}

If we could map, at fixed time, the instantaneous locations in
three-dimensional space of all Kuiper belt objects (KBOs),
on what two-dimensional surface would the density of KBOs be greatest?
We call this surface the plane
of the Kuiper belt (KBP), though by ``plane'' we do not mean to
imply that the KBP is flat (we shall find that it is not).
The KBP depends on the mass distribution of the solar system---principally,
the orbits and masses of the giant planets. There are as
many different KBPs as there are dynamical classes of KBO, since
each class of object feels a distinct time-averaged force.
Here we study the KBP defined by
Classical KBOs: objects whose fairly circular, low inclination orbits
are not in any strong mean-motion resonance with Neptune
(see Elliot et al.~2005 for a classification scheme).

In principle, theoretical determination of the KBP would help observers
to discover new KBOs.
Conversely, by measuring differences between the
theoretical KBP and the actual KBP, we might hope to infer the presence of
solar system bodies as yet undetected
(``Planet X''; see Gaudi \& Bloom 2005 for a summary of current limits).

There is disagreement regarding the location of the KBP.
Brown \& Pan (2004) analyzed the instantaneous proper motion vectors
of hundreds of KBOs irrespective of dynamical class and concluded, with
greater than $3\sigma$ confidence, that the KBP did not coincide
with the invariable plane (IP, the plane perpendicular to the total
angular momentum vector of the solar system). They argued 
that the observed KBP was consistent instead with the forced plane given by
linear secular perturbation theory. 
We will refer
to this plane as the BvWP, after Brouwer \& van Woerkom (1950),
who developed a linear secular theory for the motions of all
eight of the major planets. Their theory, in turn, has its
origin in the Laplace-Lagrange equations (see, e.g., Murray \& Dermott 1999).
The BvWP is a warped and time-variable surface
whose properties we review in \S\ref{review}.

By contrast, Elliot et al.~(2005) found that the plane determined by
Classical KBOs that were observed over multiple epochs
was more consistent with the IP
($\lesssim 1 \sigma$ difference) than with
the BvWP ($\lesssim 2$--3$\sigma$
difference).
They listed some arguments, none conclusive,
for why the IP might be preferred over the BvWP.
The low order of the BvW theory, and its inability
to account for time variations in semi-major axes,
are causes for concern.

We seek to resolve this disagreement
using numerical orbit integrations. In \S\ref{review},
we review the linear theory and how it equates the KBP with the BvWP.
In \S\ref{answer}, numerical integrations lasting the age
of the solar system are used to reveal
the theoretical location of the KBP.
In \S\ref{obs}, we compare 
theory against current observations of the KBP.
A summary is given in \S\ref{conc}, including
a brief comment on the prospects
for detecting an unseen, outer solar system planet using the KBP.

\section{LINEAR SECULAR THEORY}
\label{review}





We study the secular evolution of Classical KBOs by
solving the Laplace-Lagrange equations of motion, which
neglect all terms higher than second
order in orbital eccentricity ($e$) and orbital
inclination ($i$). The solution is
detailed in the textbook by Murray \& Dermott (1999); we provide a summary
here.

We start by describing the motions of the planets.
Define an inclination vector \mbox{ \boldmath$i$} $\equiv(q,p)\equiv (i \cos\Omega,i
\sin\Omega)$, where $i$  and $\Omega$ equal the inclination and longitude of ascending node.
Lagrange's equations governing the inclination vector
for the $j$th planet are

\begin{equation}
 \dot {q_{j}} = -{{ \frac{1}{n_j a_j^2}}{\frac{\partial
   R_j}{\partial p_j}}}
\end{equation}
\begin{equation}
\dot {p_{ j}} =  {{ \frac{1}{n_j a_j^2}}{\frac{\partial
      R_j}{\partial q_j}}} \,,
\end{equation}    
where $n_j$, $a_j$, and $R_j$ are, respectively, the mean motion, semi-major axis, and
disturbing function. We consider only the
four giant planets so that $j=1,2,3,4$ represents Jupiter, Saturn,
Uranus, and Neptune, respectively. The disturbing function, keeping
only the leading
terms relevant to the inclination evolution of the $j$th planet,
reads
\[ R_j= -\frac{n_j^2a_j^2}{8} \sum_{k=1,\neq j } ^4 \frac{m_k}{M_{\odot}+m_j} \alpha_{jk} \bar{\alpha}_{jk}
b_{3/2}^{(1)}(\alpha_{jk})\left [ q_j^2 + p_j^2 -2(q_jq_k+p_jp_k) \right ] ,\]
where $b_{3/2}^{(1)}(\alpha _{jk})$ is a 
Laplace coefficient,\footnote{
If we declare $\alpha_{jk} = \bar{\alpha}_{jk} = a_j/a_k$
and allow the Laplace coefficient to take $\alpha_{jk} > 1$ as
an argument, then separating the case $a_j < a_k$ from $a_j > a_k$ is 
not necessary. We stick here with the textbook convention, however.}
$m_j$ is the mass of the $j$th planet,
             \[\alpha _{jk} = \left\{ \begin{array}{ll} 
                             a_k/a_j & \mbox{if $a_j > a_k$}\\
                             a_j/a_k & \mbox{if $a_j < a_k$}\\
                             \end{array}
                   \right. \]
and
 \[  \bar{\alpha}_{jk} = \left\{ \begin{array}{ll}
                             1 & \mbox{if $a_j > a_k$} \\
                             a_j/a_k & \mbox{if $a_j < a_k$}. \\
                             \end{array}
                     \right.
  \]
Eqns. (1)--(2) yield
two coupled systems of first-order differential equations:
$(\dot{p_1},\dot{p_2},\dot{p_3},\dot{p_4})^T = A (q_1,q_2,q_3,q_4)^T$ 
and $(\dot{q_1},\dot{q_2},\dot{q_3},\dot{q_4})^T=
-A(p_1,p_2,p_3,p_4)^T$, where $A$ is a $4\times 4$ matrix of
constant coefficients depending on the masses and semi-major axes
of the planets. These equations describe coupled harmonic oscillators;
their solution is
\begin{equation}
q_j= \sum_{k=1}^4 I_{jk} \cos(f_kt + \gamma _k)
\label{q_j}
\end{equation}
\begin{equation}
p_j=\sum_{k=1} ^4 I_{jk} \sin(f_kt + \gamma _k) \,,
\label{p_j}
\end{equation}
where the frequencies $f_k$ are the eigenvalues of $A$. The elements
$I_{jk}$ of the eigenvectors of $A$, and the phases $\gamma_k$, are
fitted to the initial inclinations and ascending nodes of the
planets. We take initial conditions and planetary data from the NASA
JPL Horizons database for JD $=2451544.5$ (Jan 1 2000;
$t=0$ in equations (3) and
(4)). Table \ref{table_first}
lists the resultant values for $f_k$, $I_{jk}$, and
$\gamma_k$. We refer to this solution as the BvW solution,
even though Brouwer and van Woerkom (1950) included all eight planets.

\begin{deluxetable}{rrrrr}
\tablecaption{BvW solution$^{a}$ for Jan 1, 2000 ($t=0$)}
\tablewidth{0pt}

\tablehead{          & $k=1$  & 2 & 3& 4 }
\startdata
   $j=1$    &1.5792 &   0.36081  &  0.055195 &  -0.066653 \\
    2      & 1.5792 &  -0.89904  &  0.044853 &  -0.064203 \\
    3     &  1.5792 &   0.040670 & -1.01200  &  0.062630 \\
    4     &  1.5792 &   0.0045122 &  0.11941 &   0.67290    \\   \hline
   $f_{k} $ &   0 &     -123.13 &    -14.102 &    -3.2984 \\
   $ \gamma_{k}$  & 1.8788 & -0.91730 & 2.3699 & -2.7418
\enddata 
\tablenotetext{a} {The components $I_{jk}$ of the eigenvectors (in degrees),
  eigenfrequencies $f_{k}$ (in radians per Myr), and
  phases $\gamma_{k}$ (in radians), calculated using data for the giant planets
  on JD = 2451544.5 from NASA JPL Horizons. All elements are heliocentric
  and referred to the ecliptic and mean equinox of J2000.}
\label{table_first}
\end{deluxetable}

 We now turn to our main concern, the motion of a KBO of negligible mass with
 semi-major axis $a>a_j$. Its disturbing function is
\[ R= - \frac{n^2a^2}{8}\sum_{j=1}^4 \frac{m_j}{M_{\odot}} \frac{a_j}{a}
 b^{(1)}_{3/2}({a_j/a}) [q^2+p^2-2(qq_j + pp_j)] \,,\]
where unsubscripted variables refer to the KBO. Because all masses
and semi-major axes are fixed for this secular problem,
Lagrange's equations of motion are of the form
\[ \dot{q} = -\frac{1}{na^2} \frac{\partial R}{\partial p} = - c_0 p + \sum_{j=1}^4 c_jp_j \]
\[ \dot{p} = \frac{1}{na^2} \frac{\partial R}{\partial q} = + c_0 q - \sum_{j=1}^4 c_jq_j \,, \]
where the $c$'s are constants.
Substituting each equation into the time derivative of the other, we find
\begin{equation}
\ddot{q} = -c_0^2 q + \sum_{j=1}^4 c_j[c_0 q_j(t) + \dot{p}_j(t)]
\label{ddotq}
\end{equation}
\begin{equation}
\ddot{p} = -c_0^2 p + \sum_{j=1}^4 c_j[c_0 p_j(t) - \dot{q}_j(t)] \,.
\label{ddotp}
\end{equation}
Eqns.~(\ref{ddotq})--(\ref{ddotp}) describe harmonic oscillators
of natural frequency $c_0$, forced by the planetary terms in the sums.
The motion is composed of a forced oscillation and a free oscillation:
\begin{equation}{q = q_{\rm forced}+q_{\rm free}= q_{\rm forced} + i_{\rm free}\cos (ft +
 \gamma)}\label{qfree}\end{equation}
\begin{equation}
 {p = p_{\rm forced}+p_{\rm free} = p_{\rm forced} +  i_{ \rm free}\sin (ft + \gamma)} \,,\label{pfree}
\end{equation}
where $i_{\rm free}$ and $\gamma$ are constants determined by initial
conditions, and the free precession frequency
\[f = c_0 = \sum_{j=1}^4 c_j = -\frac{n}{4}\sum_{j=1}^4\frac{m_j}{M_{\odot}}\frac{a_j}{a} b_{3/2}^{(1)}(a_j/a) \,.\]
The functions $q_{\rm forced}$ and $p_{\rm forced}$ depend only on
planetary parameters and the KBO semi-major axis $a$:
\begin{equation}\label{eq_qforced}
{q_{\rm forced}=  - \sum_{k=1}^4 {{\frac{\mu_k}{f-f_k}}\cos(f_kt+\gamma_k)}}
\end{equation}
\begin{equation}\label{eq_pforced}
{p_{\rm forced}=- \sum_{k=1}^4 {{\frac{\mu_k}{f-f_k}}\sin(f_kt+\gamma_k)}}
\end{equation}
where
\[ \mu_k= -\sum_{j=1}^4 c_jI_{jk} = \frac{n}{4} \sum_{j=1}^{4} I_{jk}\frac{m_j}{M_{\odot}}\frac{a_j}{a} b_{3/2}^{(1)}(a_j/a) \,.\]

The inclination vector
\mbox{\boldmath{$i$}} of a KBO is the vector sum of a forced inclination vector \mbox{\boldmath{$i$}}$_{\rm forced} \equiv (q_{\rm forced}, p_{\rm
  forced})$ and a free inclination vector \mbox{\boldmath{$i$}}$_{\rm free} \equiv(q_{\rm free}, p_{\rm free})$. Throughout this paper, we refer to a forced inclination $i_{\rm forced}\equiv 
(q_{\rm forced}^2 + p_{\rm forced}^2)^{1/2}$,  a forced node
$\Omega_{\rm forced} \equiv \mbox{arctan} (  p_{\rm forced} /q_{\rm forced} )$, a free inclination
$i_{\rm free}\equiv (q_{\rm free}^2 + p_{\rm free}^2)^{1/2} $, and a free node $\Omega_{\rm free}\equiv \mbox{arctan} (
  p_{\rm free} / q_{\rm free} ).$ Thus free and forced inclinations
refer to magnitudes (not vectors).
The free inclination is constant; it is the undamped
amplitude of the free oscillation.

\begin{figure}
\epsscale{1.3}  
\plotone{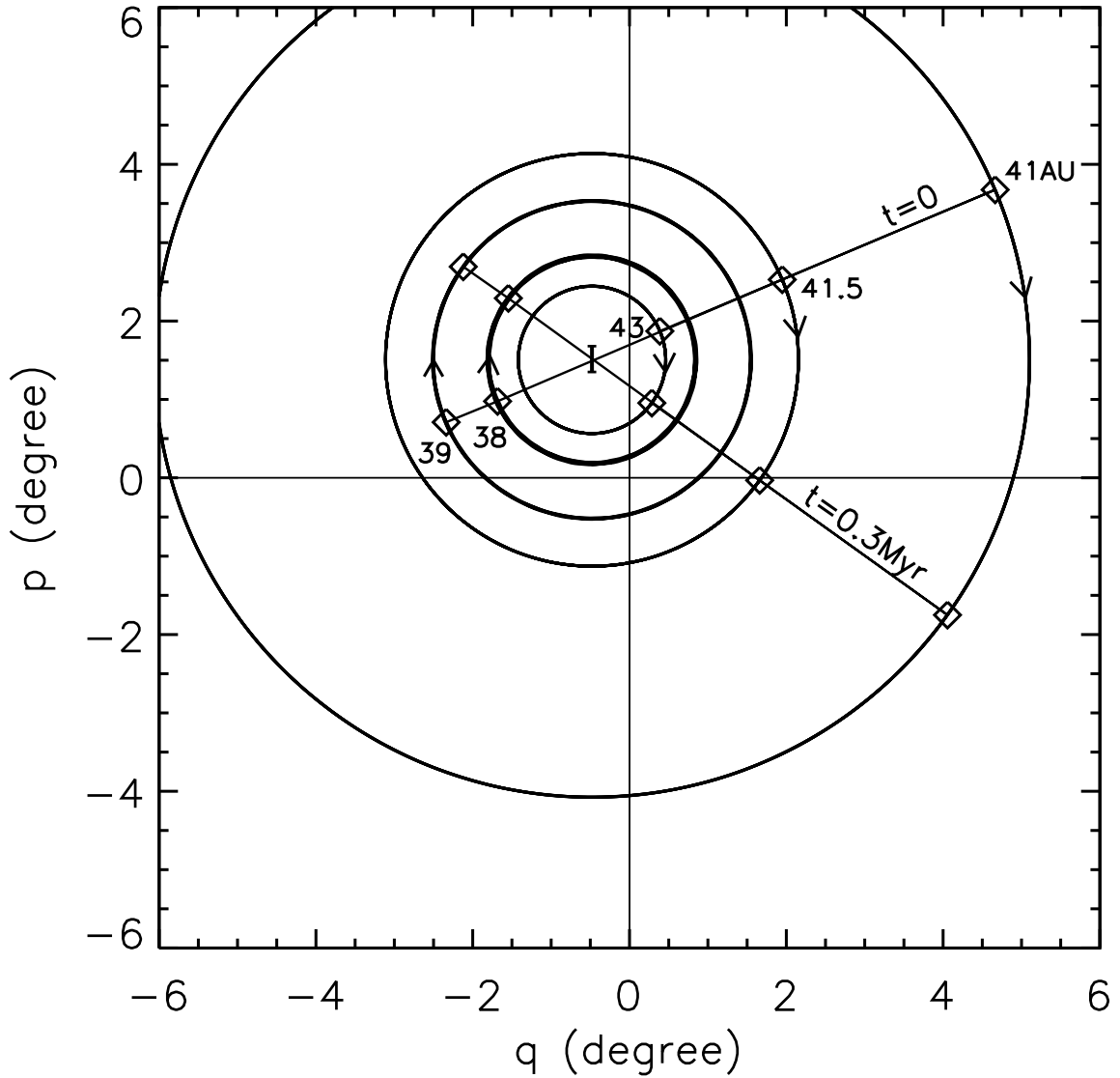}  
\caption{Locations of forced poles (open diamonds) at various semi-major axes,
at $t = 0$ (JD = 2451544.5) and $t = 0.3$ Myr. Coordinates
are referenced to the ecliptic and mean equinox of J2000 (the origin
marks the location of the J2000 ecliptic pole). The invariable pole is marked
by ``I.'' Note how the forced poles at $a > 40.5$ AU lie diametrically
opposite to those at $a < 40.5$ AU, reflecting forcing by the $\nu_{18}$
resonance at $a = 40.5$ AU. As $a$ approaches 40.5 AU, the forced pole
diverges from the invariable pole. As $a$ increases
beyond 40.5 AU, the forced pole approaches
the invariable pole. We show in this paper that the
forced poles point normal to the Kuiper belt plane.
}  
\label{fig_diameter}
\end{figure}

\placefigure{fig_diameter}

Figure \ref{fig_diameter}
shows the evolution of \mbox{\boldmath{$i$}}$_{\rm forced}$
at various semi-major axes in the Kuiper belt. The open diamonds,
located at the endpoints of \mbox{\boldmath{$i$}}$_{\rm forced}$,
mark what we call the forced poles. At fixed time, the forced poles
all lie along a line intersecting the invariable
pole, denoted \textbf{I}. To the extent that the $k=4$ mode, driven
mainly by Neptune, dominates, the forced
poles rotate clockwise (regress) about the invariable pole at a single
frequency ($f_4$), maintaining constant distance to \textbf{I}.
Note that the forced poles at
$a>40.5$AU lie diametrically opposite to those at $a<40.5$AU, and that
as $a$ approaches 40.5AU from either above or below, the separation
between the forced pole and \textbf{I} increases. These
latter two properties reflect the $\nu_{18}$ secular resonance at
$a=40.5$AU, where the denominators $f-f_4$ of equations
(\ref{eq_qforced})--(\ref{eq_pforced}) vanish:
the forced response becomes infinite in magnitude at resonance,
and changes phase by $180^{\circ}$ across resonance (the sign
of $f-f_4$ switches across $a = 40.5$AU).

\begin{figure}  
\epsscale{1.2}  
\plotone{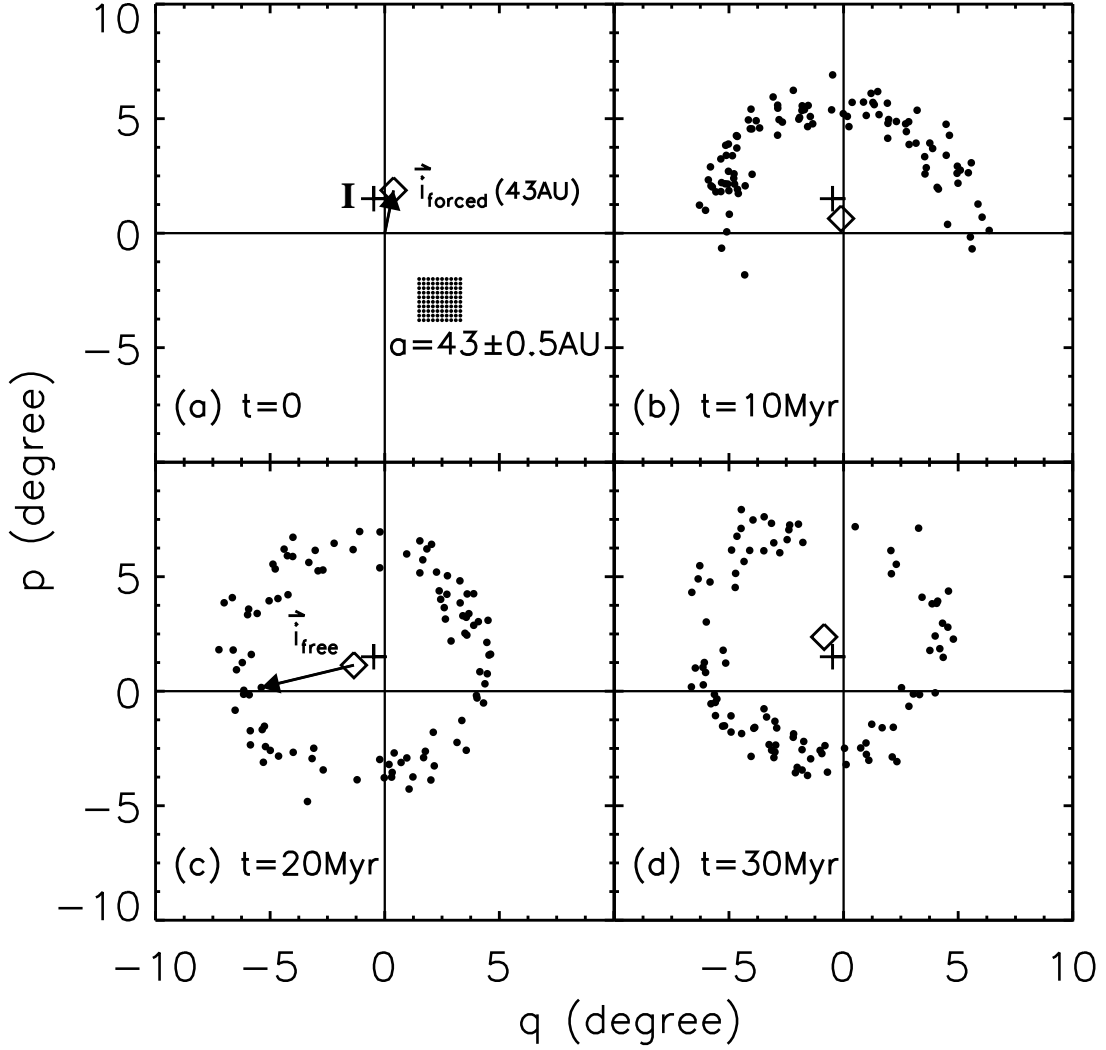}  
\caption{ 
Phase mixing and axisymmetry of KBOs about the forced pole.
In panel (a), the orbit poles of 100 test particles (solid circles),
having semi-major axes randomly distributed between 42.5 and 43.5 AU,
are distributed on a grid in $p$-$q$ space.
The forced pole at $a = 43$ AU (open diamond)
and the invariable pole (``+'') are indicated. As time
elapses from panels (b) through (d), the poles of the test particles
precess about the forced pole, which itself precesses about the invariable
pole. Precession rates differ from one particle to the next
according to their semi-major axes, while the distance from
each pole to the forced pole---the free inclination
(see panel c)---remains constant. Consequently, after a few tens of Myrs,
orbit poles tend to phase mix into an annulus centered on the forced pole
(not the invariable pole). The evolution shown here was computed
using the analytic BvW solution, not the numerical integration.
}  
\label{fig_phasemix}
\end{figure}

\placefigure{fig_phasemix}

Figure \ref{fig_phasemix} illustrates how free nodes $\Omega_{\rm free}$
phase mix and how
such phase mixing helps determine the KBP within the BvW theory.
At $t=0$, one hundred KBOs having semi-major axes within 0.5
AU of 43 AU are set down with orbit normals approximately aligned
about an arbitrary direction
(the orbit poles are actually distributed in a small box
in $p$-$q$ space). Over tens of Myrs, the orbit poles of the particles
drift away from one another: the small dispersion
in semi-major axis produces a small dispersion
in the free precession frequency $f$ (the rate at which the
free inclination vector rotates about the forced pole).\footnote{In the linear
theory, precession frequencies $f$ and $f_k$ do not depend on
eccentricity or inclination, but in higher order theories they do.}
While the free nodes
distribute themselves over all phases, the free inclinations $i_{\rm free}$
remain fixed.
Thus, at late times, the collection of
free inclination vectors \mbox{\boldmath{$i$}}$_{\rm free}$
are distributed
axisymmetrically about the mean forced pole at $a \approx 43$ AU.
The axisymmetry arises practically independently of how the particles'
orbit normals are initially distributed; the only requirements
are that many particles share the same $i_{\rm free}$ (so that
such particles, when phase mixed, trace a full circle) and that
there exists a small but non-zero dispersion in semi-major axis
(so that there exists a non-zero spread in free
precession frequencies, enabling phase mixing).
Note in Figure \ref{fig_phasemix}
how the invariable pole does not coincide
with the center of the circular distribution.

We conclude that according to the BvW theory,
the mean orbit normal of a phase-mixed group
of particles having approximately the same semi-major axis
is given by the forced pole corresponding to that semi-major axis.
That local forced pole varies
with time (Figure \ref{fig_diameter}),
but the particles always encircle it (Figure \ref{fig_phasemix}).
The KBP is warped because the forced pole changes direction
with KBO semi-major axis (Figure \ref{fig_diameter}).

Note finally that the ability of test particles to keep their free
inclinations constant relative to a time-variable forced pole 
should not be confused with adiabatic invariance.
Generally the frequencies $f$ and $f_k$ are not cleanly separated.
The constant $i_{\rm free}$ simply reflects the undamped amplitude
of free oscillation, and is set by initial conditions. 
In the next section, we use numerical simulations
to test the constancy of free inclination.

\section{NUMERICAL INTEGRATIONS}
\label{answer}

\subsection{Initial Conditions}
\label{ic}

We calculate the evolution of the KBP at three semi-major axes:
$a = 38$, 43, and 44 AU. These are chosen to lie away
from strong mean-motion resonances (e.g., the 3:2 resonance resides at
39.5 AU) and outside the $a = 40$--42 AU region of instability
carved by the $\nu_{18}$, $\nu_{17}$ and $\nu_8$ secular
resonances (see, e.g., Chiang et al.~2007). At each $a$
we lay down $N_i \times N_{\Omega}$ test particles
whose initial free inclination vectors are distributed axisymmetrically
about the local forced pole. That is, each particle's initial
$p = p_{\rm forced}(a) + p_{\rm free}$ and
initial $q = q_{\rm forced}(a) + q_{\rm free}$, where
$i_{\rm free} = (p_{\rm free}^2 + q_{\rm free}^2)^{1/2}$ takes
1 of $N_i=4$ values (0.01, 0.03, 0.1, 0.3 rad) and
$\Omega_{\rm free} = \arctan (p_{\rm free}/q_{\rm free})$
takes 1 of $N_{\Omega} = 20$ values distributed uniformly
between 0 and $2\pi$. This set-up permits us to
directly test the BvW theory, which predicts that
all $N_{\Omega}$ particles corresponding to a given $a$
and given initial $i_{\rm free}$ should keep the
same $i_{\rm free}$ for all time: a circle of points
in $p$-$q$ space should continue to trace the same-sized
circle. Coordinates $p_{\rm forced}$
and $q_{\rm forced}$ for the initial forced poles are
computed using the BvW solution of \S\ref{review}, for $t = 0$.

Initial osculating eccentricities are zero and initial mean anomalies
are chosen randomly between 0 and $2\pi$. The four giant
planets are included in the integration, with initial
conditions taken from the JPL Horizons database for
JD = 2451544.5 ($t=0$ in the BvW theory).
The integration is performed with the swift\_rmvs3
code, written by Levison \& Duncan (1994) and based
on the algorithm developed by Wisdom \& Holman (1991).
The duration of the integration is 4 Gyr and the timestep
is 400 days (about 1/11 the orbital period of Jupiter).
We work in a heliocentric coordinate system, the better to compare with
the linear secular theory which uses heliocentric elements.

\subsection{Results}
\label{res}

\begin{figure}  
\epsscale{1.2}
\plotone{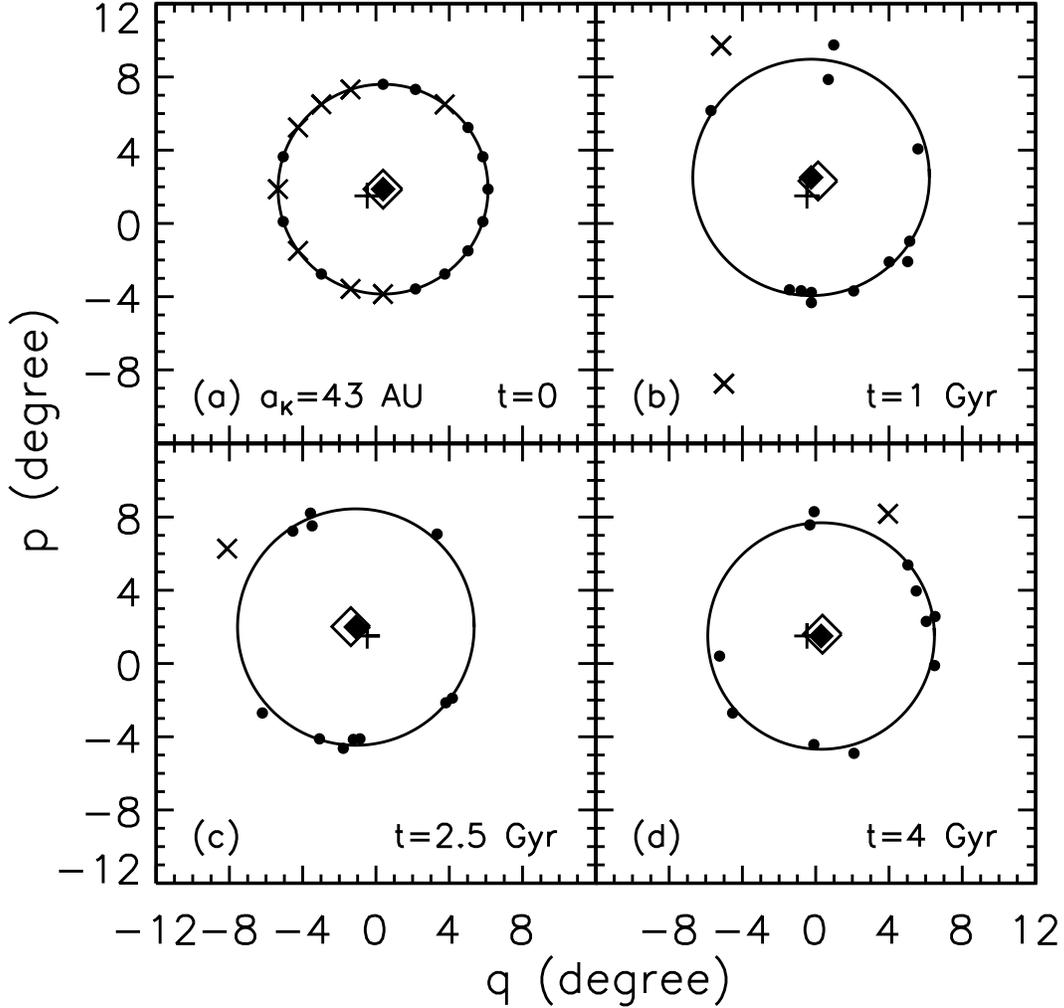}
\caption{Tracking the Kuiper belt plane by numerical integration.  In
  panel (a), we set up $N_{\Omega}=20$ test particles having initial
  semi-major axes of $a_{\rm K}=43$ AU and having orbit poles
  distributed in a cone of half-width $i_{\rm free}=0.1$ rad centered on the
  local forced pole given by the BvW solution.  In panels (b) through
  (d), the orbit poles evolve according to our numerical
  integration. Solid circles denote test particles that both survive
  the entire 4 Gyr duration of the integration and have osculating
  eccentricities less than $e_{\rm cut} = 0.08$; these are fitted to a
  circle, whose center yields the local normal to the Kuiper belt
  plane (solid diamond). The ``X'' symbols denote test particles that
  do not satisfy these requirements. The Kuiper belt pole so obtained
  follows closely that predicted by the continuously updated BvW
  solution (open diamond), and does not point along the invariable
  pole (``+'').  }
\label{43_0.10}  
\end{figure}

According to the BvW theory, each set of $N_{\Omega} = 20$
particles having the same initial $a$ and initial
$i_{\rm free}$ should trace the perimeter of a single circle
in $p$-$q$ space at any given time, with the
center of the circle yielding the local normal to the KBP.
Figure \ref{43_0.10} tests this prediction; the panels
display the $p$-$q$ positions of particles initially having $a = 43$ AU
and $i_{\rm free} = 0.1$ rad, sampled at four different times.
Most particles at a given time do lie approximately on one circle, although
there are outliers (see crosses in panels b, c, and d).
The outliers represent particles whose inclinations and eccentricities
grow to large values. Many of these particles suffer
close encounters with Neptune, whereupon they are removed from
the integration.
We did not identify the cause of the instability, but probably the
various high-order mean-motion resonances in the vicinity (Nesvorn\'y \&
Roig 2001) are to blame.
Of the 20 particles shown at $t=0$ in panel a, only
12 survive to $t = 4$ Gyr in panel d. Those that survive have
semi-major axes that remain constant to within $\pm$0.5 AU.

\placefigure{fig_stir}

At each $t$, we fit a circle (in a least squares sense) to those
particles known to survive the entire length of the integration.
The fit is substantially improved by also discarding particles
whose instantaneous eccentricities exceed some value $e_{\rm cut}$, as we find
that particles on eccentric orbits tend also to be outliers in $p$-$q$ space.
The value of $e_{\rm cut}$ is reduced from one until the fit parameters
cease to change significantly. Note that a particle that is discarded
from the fit by the $e_{\rm cut}$-criterion at one
time can be restored to the fit at a later time (if its eccentricity
falls below $e_{\rm cut}$ at that later time).
Table \ref{stability} lists the values of $e_{\rm cut}$
chosen for the various combinations of initial $a$ and initial $i_{\rm free}$.

\begin{figure}  
\epsscale{1.2}  
\plotone{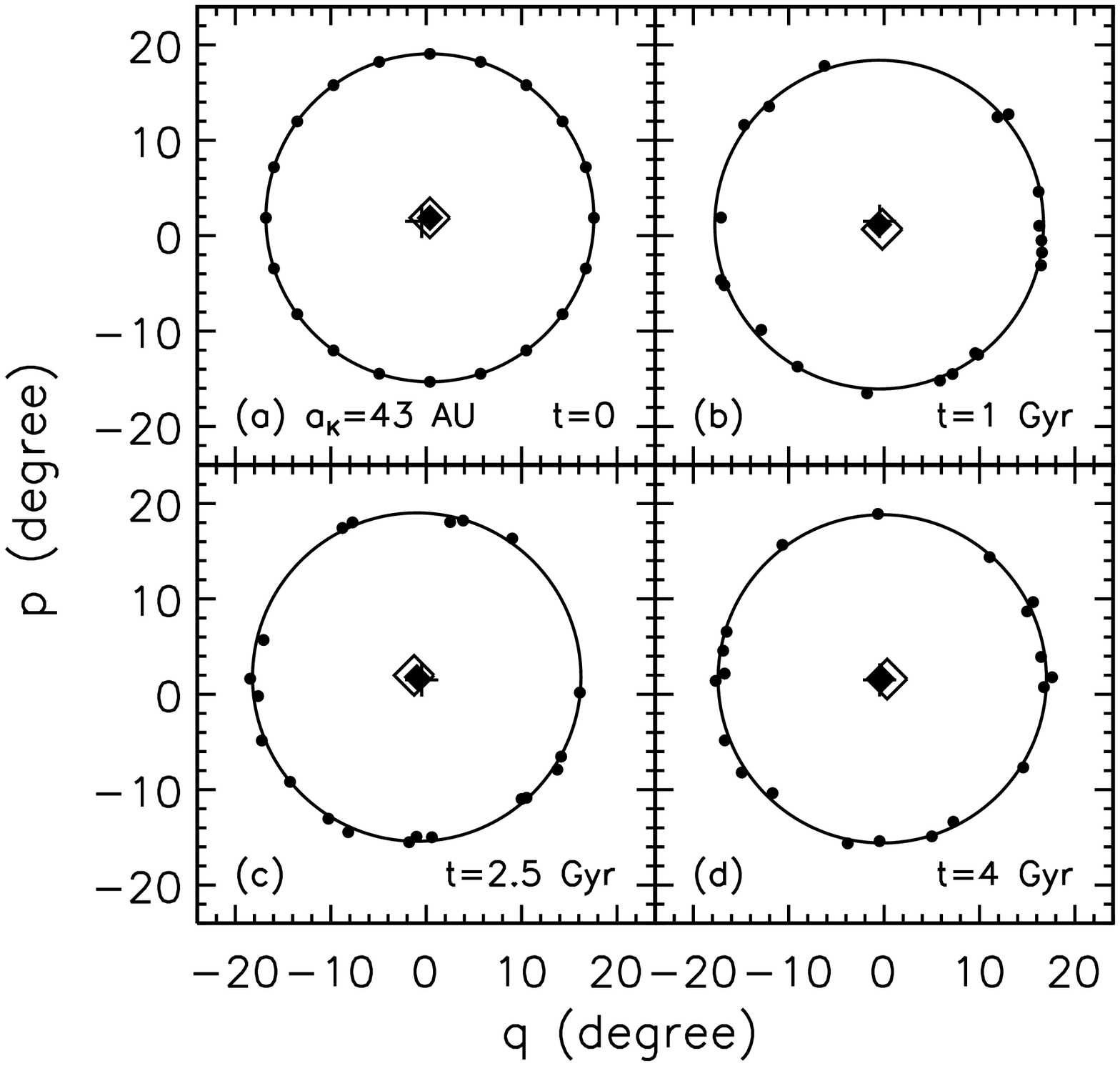}  
\caption{Same as Figure \ref{43_0.10}, except that
initial $a_{\rm K}=43$ AU, initial $i_{\rm free} = 0.3$ rad,
and no $e_{\rm cut}$ criterion is applied. The numerically
fitted Kuiper belt pole (solid diamond) tracks the continuously
updated BvW pole (open diamond) well; both precess about the invariable pole
(``+'').
} 
\label{43_0.30}  
\end{figure}

\begin{figure}  
\epsscale{1.2}  
\plotone{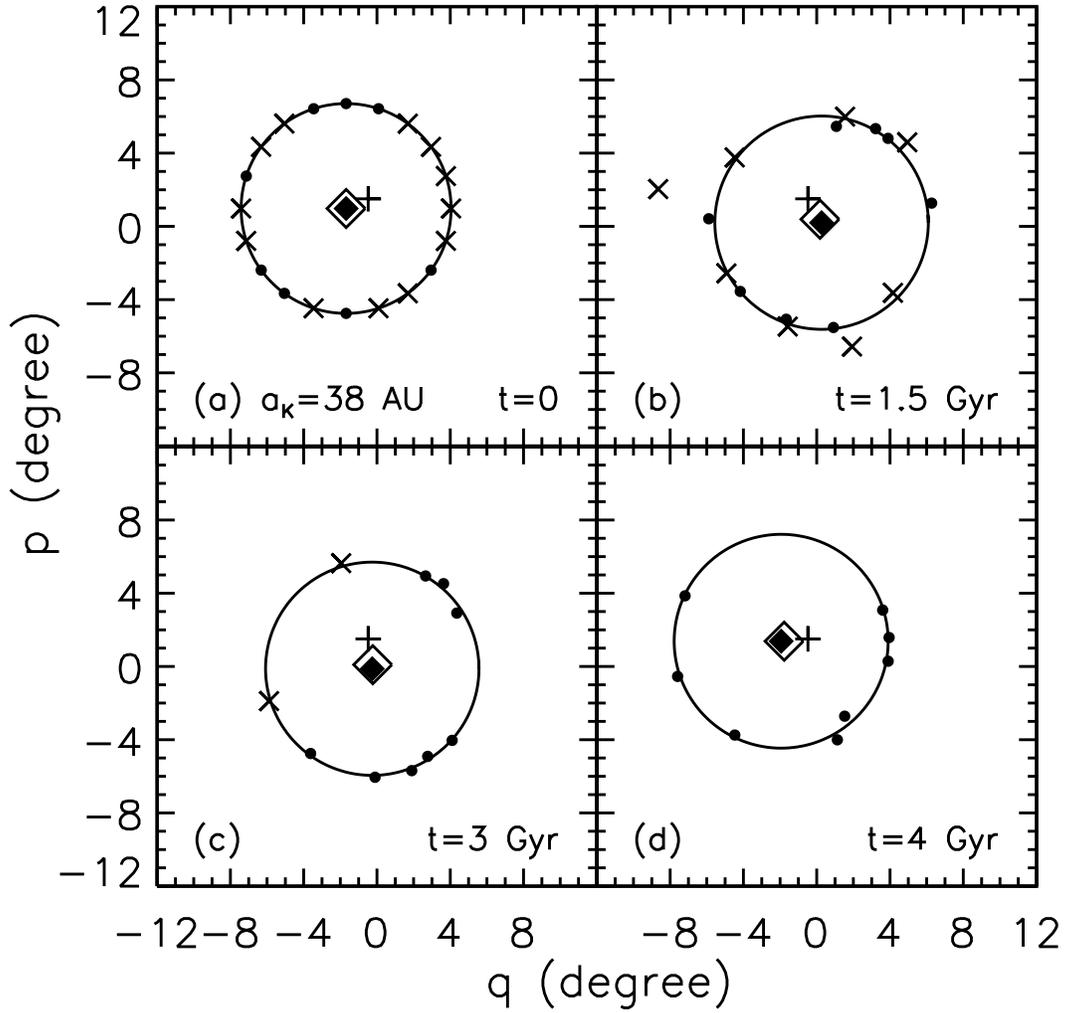}  
\caption{Same as Figure \ref{43_0.10}, except that
initial $a_{\rm K}=38$ AU, initial $i_{\rm free} = 0.1$ rad,
and no $e_{\rm cut}$ criterion is applied. 
Here again
the numerically determined Kuiper belt pole (solid diamond)
is well described by the continuously updated BvW pole (open diamond),
not the invariable pole (``+'').
}  
\label{38_0.10}  
\end{figure}

\begin{figure}  
\epsscale{1.2}  
\plotone{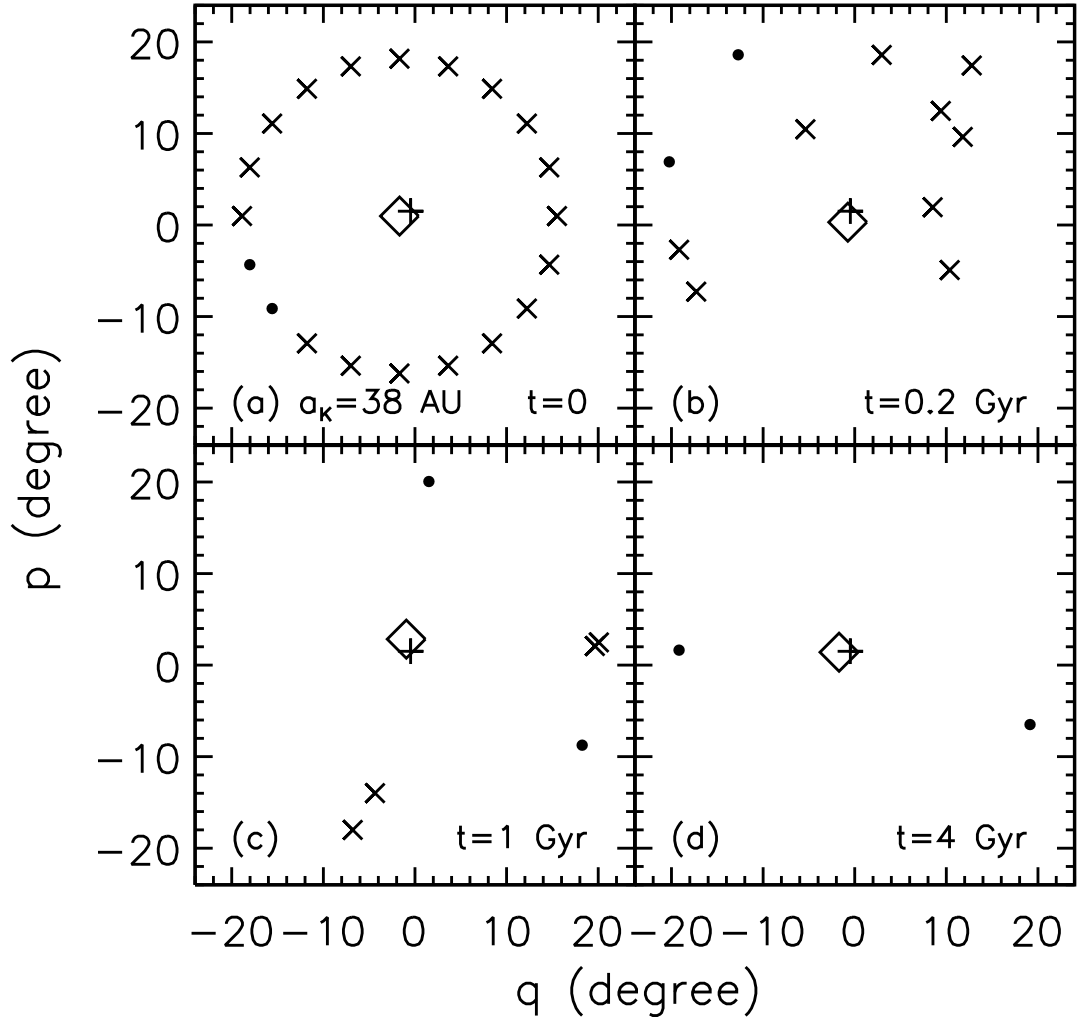}  
\caption{Same as Figure \ref{43_0.10}, except that
initial $a_{\rm K}=38$ AU, initial $i_{\rm free} = 0.3$ rad,
and no $e_{\rm cut}$ criterion is applied. 
Only two test particles survive the 4 Gyr duration of the
integration, rendering determination of the Kuiper belt pole impossible.
}  
\label{38_0.30}  
\end{figure}

The circles so fitted are overlaid in Figure \ref{43_0.10}. A typical
fit is excellent and the local normal of the KBP (center of the fitted circle;
solid diamond) is confidently identified.
Figures \ref{43_0.30}, \ref{38_0.10}, and \ref{38_0.30} report
analogous results for
other choices of initial $a$ and initial $i_{\rm free}$. All 20 particles
having $a = 43$ AU and $i_{\rm free} = 0.3$ rad are stable for 4 Gyr
(Figure \ref{43_0.30}). By contrast, at $a=38$ AU,
where various high-order mean-motion resonances are known
to cause instability (Nesvorn\'y \& Roig 2001),
only 2 out of 20 particles having $i_{\rm free} = 0.3$ rad
remain at the end of 4 Gyr (Figure \ref{38_0.30}).
Fitting a unique circle to 2 points is impossible. And
as is clear from Figure \ref{38_0.30},
even if we were to try fitting circles at earlier times when more particles
are present, such fits would be poor.
Only certain combinations of $a$-$i_{\rm free}$ permit
determination of the KBP, as documented in our Table \ref{stability}.

\begin{figure}  
\epsscale{1}
\plotone{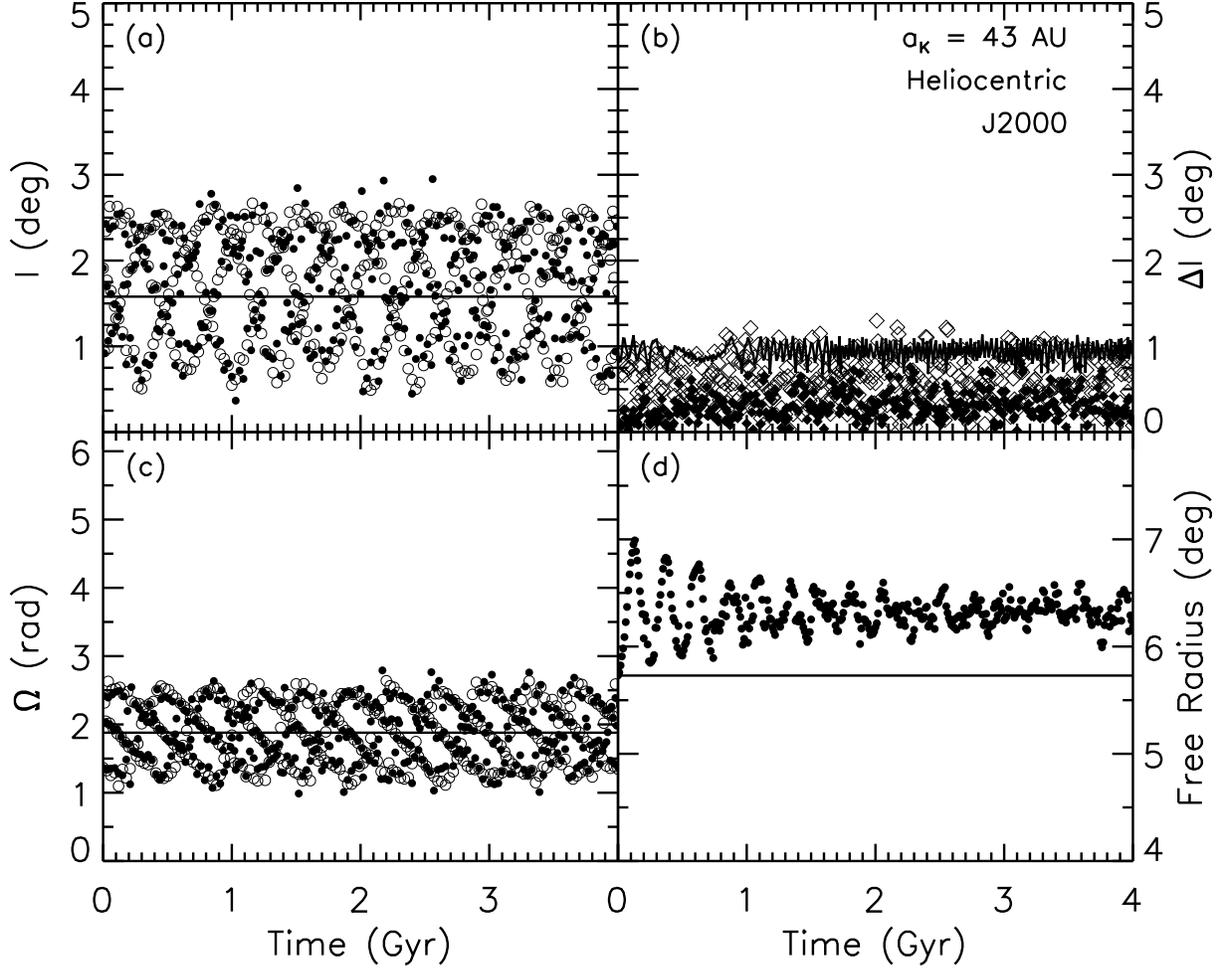}  
\caption{Results for test particles
having initial $a_{\rm K} = 43$ AU and initial $i_{\rm free}=0.1$ rad,
demonstrating that the numerically obtained KBP follows the continuously
updated BvWP, not the IP.
(a) Inclination of the KBP (solid circles),
BvWP (open circles), and IP (line), all relative
to the ecliptic. (b) Mutual inclination between the KBP and BvWP (solid
diamonds), the BvWP and the IP (open diamonds), and the KBP and the IP
(solid line). (c) Longitude of ascending node of the KBP (solid circles),
BvWP (open circles), and IP (line), all relative to the ecliptic
and mean equinox of J2000. (d) Numerically obtained free inclination
(solid circles; these are the radii of the fitted circles in
Figure \ref{43_0.10}), compared against the initial $i_{\rm free}$
(line). The BvW theory predicts that the free inclination should be constant;
while it is nearly so in our numerical integration, its mean value
is offset by 0.6 deg (10\%) relative to its initial value.
Data are sampled every $10^8$ yr and do not resolve
the precession of the KBP occurring with a $\sim$$10^6$-yr period
(see Figure \ref{fig_ad}).
}  
\label{fig_fourplot_a}  
\end{figure}

\begin{figure}  
\epsscale{1}  
\plotone{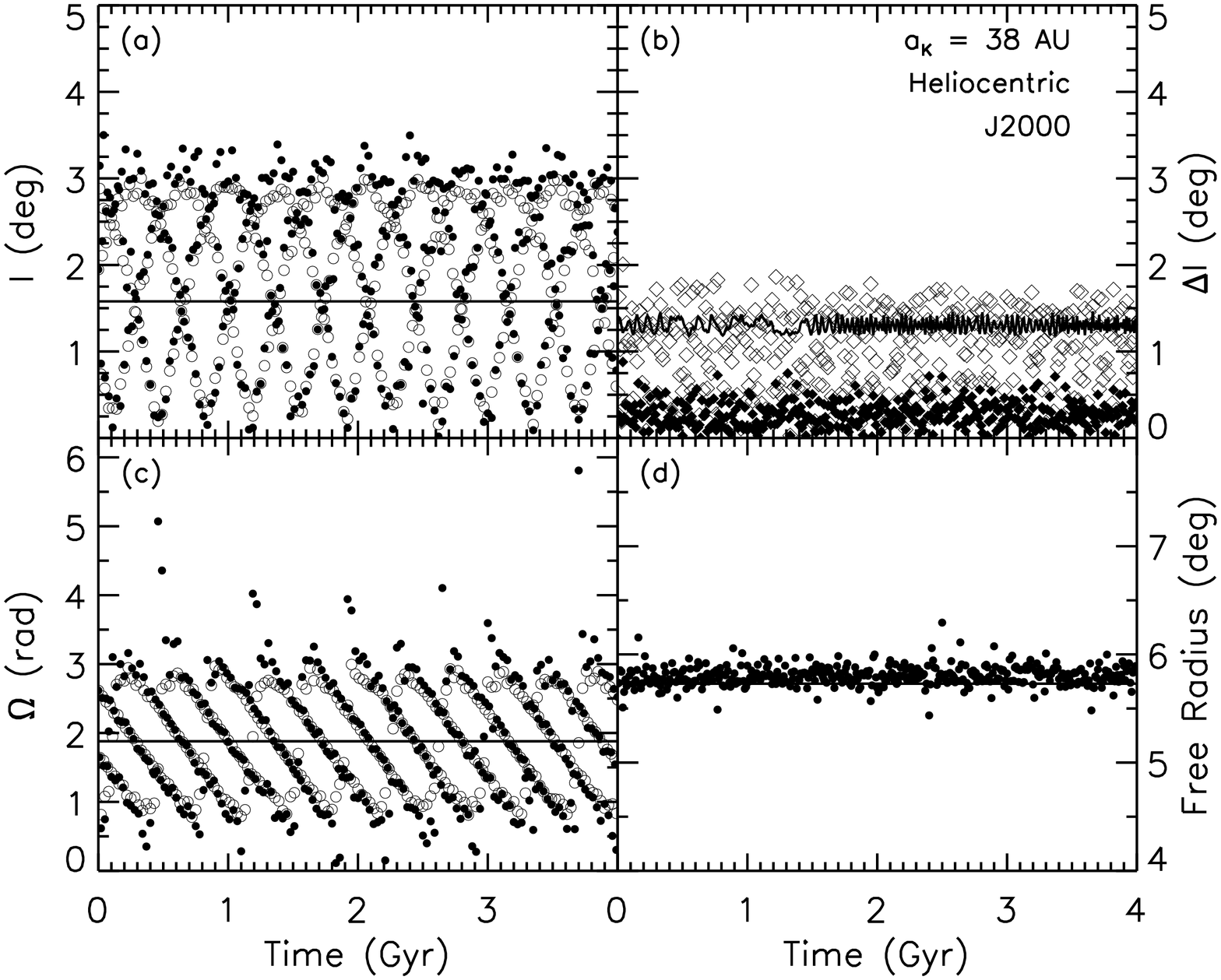}  
\caption{Same as Figure \ref{fig_fourplot_a}, except
for initial $a_{\rm K} = 38$ AU and initial $i_{\rm free} = 0.1$ rad.
}  
\label{fig_fourplot_b}  
\end{figure}

Panels d of Figures \ref{fig_fourplot_a} and \ref{fig_fourplot_b}
plot the radii of the fitted circles versus time, for two
choices of initial $a$-$i_{\rm free}$. The BvW linear theory predicts
that these radii should be constant. In fact they are nearly so,
varying by at most one part out of six.

How can we further test the BvW theory when we know that it fails to
predict the orbits of the planets on Gyr timescales? We compute
instead a semi-analytic, BvW-based solution as follows. At each time
in the numerical integration, we output the inclinations and ascending nodes of
the giant planets and use these to recompute the eigenvectors---and
thus the forced poles---of the linear theory.  Thus we obtain a
prediction of where the KBP should reside according to the linear
theory at a given instant, using the simulation results for the planets at that
instant to supply the integration constants. This ``continuously
updated BvW solution'' for the local normal (forced inclination vector) is shown
as an open diamond in Figures \ref{43_0.10}--\ref{38_0.30}.
Its location tracks that
of the numerically fitted pole (solid diamond) well---much better than
does the invariable pole (upright cross); see especially Figures
\ref{43_0.10} and \ref{38_0.10}.
Figures \ref{fig_fourplot_a} and \ref{fig_fourplot_b} also demonstrate
that the continuously updated BvWP hews closely
to the numerically fitted KBP.

\placefigure{fig_four}

\placefigure{fig_five}

\placefigure{fig_six}

\begin{deluxetable}{cccccc}
\tablecaption{Numerical Results for Stability in Initial $a$-$i_{\rm free}$ Space}
\tablewidth{0pt}
\tablehead{
\multicolumn{6}{c}{}\\
$a$ & $i_{\rm free}$ & \# of Survivors & $e_{\rm max}$ & $e_{\rm cut}$ & Permits KBP \\
(AU) & (rad) & (out of 20) & of Survivors & & Determination? \\
}
\startdata
38 & 0.01 & 17 & 0.05 & None Applied & Yes \\
   & 0.03 & 16 & 0.05 & None Applied & Yes \\
   & 0.10 & 8  & 0.07 & None Applied & Yes \\
   & 0.30 & 2  & 0.02 & None Applied & No \\

43 & 0.01 & 19 & 0.12 & 0.04 & Yes \\
   & 0.03 & 20 & 0.04 & None Applied & Yes \\
   & 0.10 & 12 & 0.12 & 0.08 & Yes \\
   & 0.30 & 20 & 0.02 & None Applied & Yes \\

44 & 0.01 & 20 & 0.11 & 0.03 & Yes \\
   & 0.03 & 20 & 0.11 & 0.04 & Yes \\
   & 0.10 & 20 & 0.15 & 0.06 & Yes \\
   & 0.30 & 20 & 0.29 & 0.08 & Yes \\
\enddata
\label{stability}
\end{deluxetable}

To summarize the results of our numerical simulations:
Linear secular theories like BvW correctly predict
how the warped KBP evolves with time, provided the parameters
of those analytic theories are continuously updated using
either observations or numerical simulations of the 
giant planets' orbits. The KBP today is accurately
predicted by the updated BvW solution.
The KBP does not, in general,
coincide with the invariable plane, except at infinite distance
from the planets.

\placefigure{fig_seven}

\placefigure{fig_eight}

\section{THEORY VERSUS OBSERVATION}
\label{obs}

We have shown by numerical simulations in \S\ref{answer} that the Classical KBP
is given by the BvWP, i.e., by linear secular theory.
Here we assess whether observations of KBOs bear out this result,
by locating the actual poles of the KBP near 38 AU and 43 AU.
These semi-major axes lie to either side of the $\nu_{18}$ resonance;
theory predicts that the corresponding poles should lie to either
side of the invariable pole, with all three orbit normals lying in one
plane (see Figure \ref{fig_diameter}). Our dataset consists
of KBOs listed on the Minor Planet Center
website on Jan 22 2008 whose (a) astrometric arcs extend longer than 50 days
(many objects in our sample have much longer arcs),
(b) eccentricities are less than 0.1, (c) inclinations
are less than 10 degrees, and (d) are classified by
the Deep Ecliptic Survey (DES) as ``Classical'' in at least
two out of their three orbital integrations (see Elliot et al. 2005
for a description of the classification scheme).
In the vast majority of cases, all three DES integrations
yield a classification of ``Classical.''
Moreover, the typical $3\sigma$ uncertainties in semi-major axes
are smaller than $\sim$0.1 AU.
We assemble two samples, one for which
$38.09 \,{\rm AU} < a < 39.10 \,{\rm AU}$ (the ``38 AU'' sample,
containing 10 objects)
and another for which $42.49 \,{\rm AU} < a < 43.50\,{\rm AU}$ (the
``43 AU'' sample, containing 80 objects).
Object designations are given in Table \ref{table_desig}.

\begin{deluxetable}{cl}
\tablecaption{Samples of Observed KBOs for Locating the KBP}
\tablewidth{0pt}

\tablehead{ Sample & Objects}
\startdata
   ``38 AU''    & 1998 WV$_{24}$, 1999 OJ$_4$, 2000 YB$_2$, 82157, 2003 FD$_{128}$, 2003 QA$_{92}$, 2003 YL$_{179}$, \\ 
                & 2003 QQ$_{91}$, 144897, 119951\\
   ``43 AU''    & 19255, 1994 EV$_3$, 1996 TK$_{66}$, 33001, 1998 WY$_{24}$, 1998 WX$_{24}$, 1999 CN$_{153}$,\\
                & 1999 RT$_{214}$, 1999 ON$_4$, 1999 XY$_{143}$, 1999 RW$_{214}$, 1999 CH$_{154}$, 1999 RU$_{215}$,\\
                & 1999 HV$_{11}$, 1999 DA, 1999 HJ$_{12}$, 1999 CW$_{131}$, 2000 PU$_{29}$, 2000 PX$_{29}$,\\
                & 134860, 2000 CL$_{105}$, 2000 ON$_{67}$, 2000 PC$_{30}$, 2000 FS$_{53}$, 2000 WV$_{12}$,\\
                & 2000 WL$_{183}$, 2000 OU$_{69}$, 2000 YU$_1$, 88268, 2001 QB$_{298}$, 2001 QD$_{298}$, 2001 XR$_{254}$,\\
                & 2001 QO$_{297}$, 2001 HZ$_{58}$, 2001 RW$_{143}$, 2001 OK$_{108}$, 2001 DB$_{106}$, 88267,\\
                & 2001 XU$_{254}$, 2001 FK$_{185}$, 2001 OZ$_{108}$, 2002 CD$_{251}$, 2002 PX$_{170}$, 2002 PV$_{170}$,\\
                & 2002 FW$_{36}$, 2002 WL$_{21}$, 160256, 2002 VB$_{131}$, 2002 PY$_{170}$, 2002 CS$_{154}$, 2002 PD$_{155}$,\\
                & 2003 SN$_{317}$, 2003 UT$_{291}$, 2003 FK$_{127}$, 2003 QG$_{91}$, 2003 FA$_{130}$, 2003 HY$_{56}$,\\
                & 2003 QY$_{90}$, 2003 TK$_{58}$, 2003 QF$_{91}$, 2003 QE$_{91}$, 2003 QZ$_{111}$, 2003 QL$_{91}$, 2003 QE$_{112}$,\\
                & 2003 YR$_{179}$, 2003 QY$_{111}$, 2003 QD$_{91}$, 2003 TL$_{58}$, 2003 QU$_{90}$, 2003 YT$_{179}$,\\
                & 2003 YX$_{179}$, 2003 YS$_{179}$, 2003 YJ$_{179}$, 2004 UD$_{10}$, 2004 DM$_{71}$, 2005 JZ$_{174}$,\\
                & 2005 GD$_{187}$, 2005 JP$_{179}$, 2005 XU$_{100}$, 2006 HA$_{123}$ \\
\enddata 
\label{table_desig}
\end{deluxetable}

\placefigure{fig_nine}
\begin{figure}  
\epsscale{1}  
\plotone{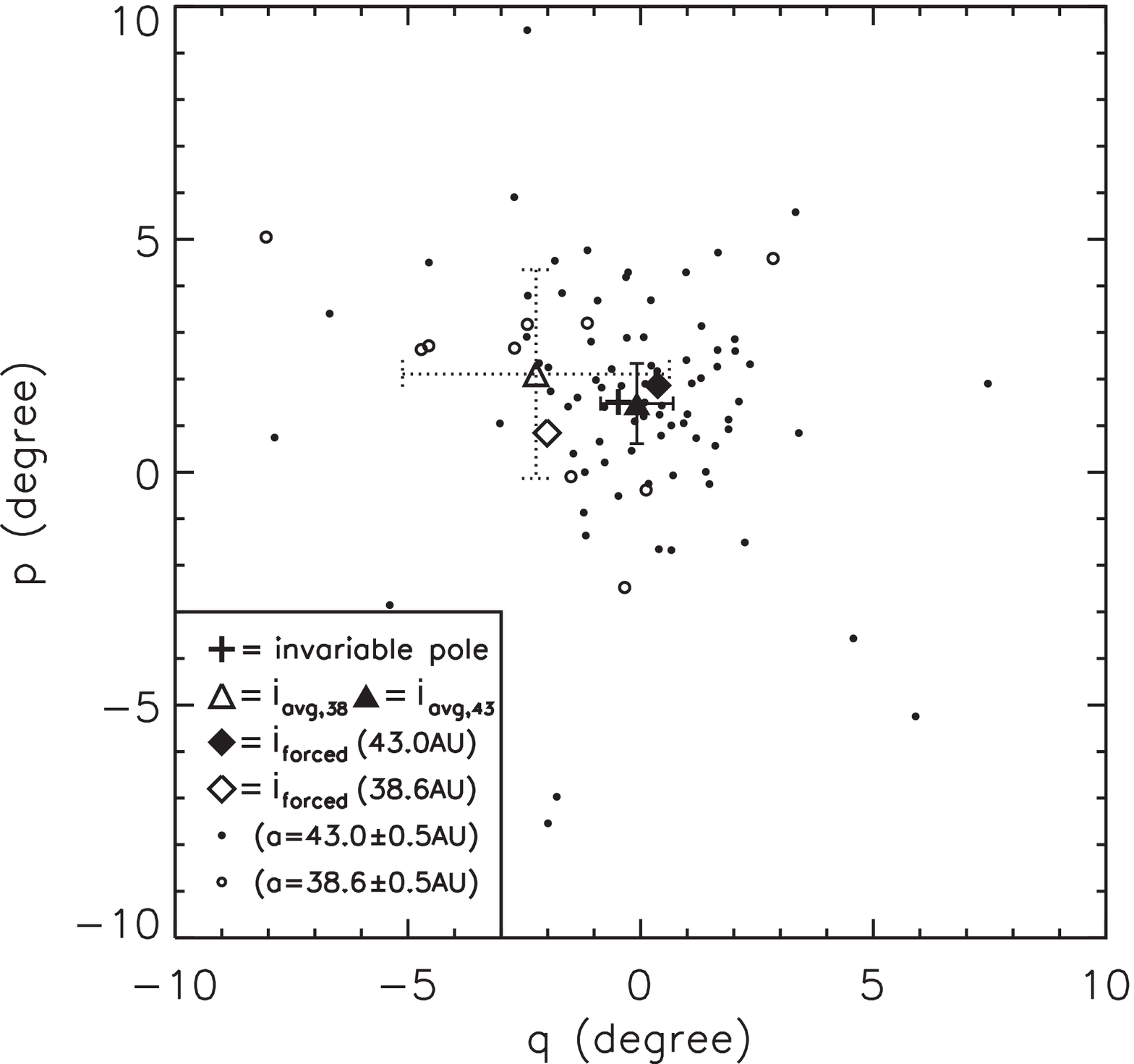}  
\caption{Pole positions of observed Classical KBOs having semi-major axes
near 38 AU (open circles) and 43 AU (filled circles), in
$q = i \cos \Omega, p = i \sin \Omega$ space, referenced
to the J2000 ecliptic. According
to theory (\S\ref{review}--\S\ref{answer}), the average pole position
at a given semi-major axis (triangles with error bars) should match the
forced pole positions (diamonds) calculated from BvW. They do
to within $2\sigma$ (error bars are $\pm 3\sigma$,
where $\sigma$ is the standard deviation of the mean).
Unfortunately, the alternate hypothesis that the average pole positions
are given by the invariable pole (bold cross) cannot be ruled out
with greater confidence.
}  
\label{fig_real}  
\end{figure}

Figure \ref{fig_real} plots the observed ($q$,$p$) positions for the two samples,
and compares with theory.
Theory predicts that the average $(\bar{q},\bar{p})$ measured for each
sample should equal $(q_{\rm forced}, p_{\rm forced})$ calculated for
the average semi-major axis of the sample. The good news is that the
data are consistent with this prediction; the differences between the
observed poles and the forced poles are less than 2$\sigma$ for both
samples (the error bars in Figure \ref{fig_real} are $\pm 3\sigma$,
where $\sigma$ is the standard deviation of the mean).  The bad news
is that the observed poles are also consistent with the invariable
pole, at a similar confidence level.  
While it is encouraging for the theory that the observed pole at 38 AU indeed
 lies to
the left (toward smaller $q$) of the invariable pole, and that the observed
pole at 43 AU lies to the right, there are not enough
observations to make more precise
statements and to rule out the hypothesis that the KBP equals the IP with
greater than $3\sigma$ confidence.

Note that by selecting our sample to have orbital inclinations less
than 10 degrees with respect to the ecliptic plane, we bias our measurement
of the average pole position towards the ecliptic pole. This systematic
error is probably still smaller, however, than our random error.
For example,
the observed pole at 38 AU actually lies further
from the ecliptic pole
than does the theoretically expected forced pole. See Elliot et al. (2005)
for ways of reducing this systematic bias.

We also performed a two-dimensional Kolmogorov-Smirnov test to see whether
the $(q,p)$ distributions for the two samples differ (they should).
The probability that they do not is 4.9\%---small enough
to be suggestive of a real difference, but in our judgement
too large to be conclusive.


\section{SUMMARY AND DISCUSSION}
\label{conc}

Classical Kuiper belt objects trace a sheet that warps and precesses
in response to the planets. The
current location and shape of this sheet---the ``plane'' of the
Classical Kuiper belt---can be computed using linear secular theory,
with the observed masses and current orbits of the planets as input
parameters (Brown \& Pan 2004).  The local normal to the plane is given by the
theory's forced inclination vector, which is
specific to every semi-major axis. At infinite
distance from the planets, the plane coincides with the invariable
plane. The deviations of the Kuiper belt
plane away from the invariable plane, while generally non-zero, are
typically small: less than 3
degrees outside the secularly unstable gap at $a \approx 40.5 \pm 1$
AU (inside the gap no KBOs have been observed, as
expected).  As the semi-major axis varies from $< 40.5$ AU to $> 40.5$ AU,
the ascending node of the Kuiper belt
plane on the invariable plane rotates by very nearly 180 degrees,
a result of the sign change in the forced response across
the $\nu_{18}$ resonance.

\begin{figure}  
\epsscale{1}  
\plotone{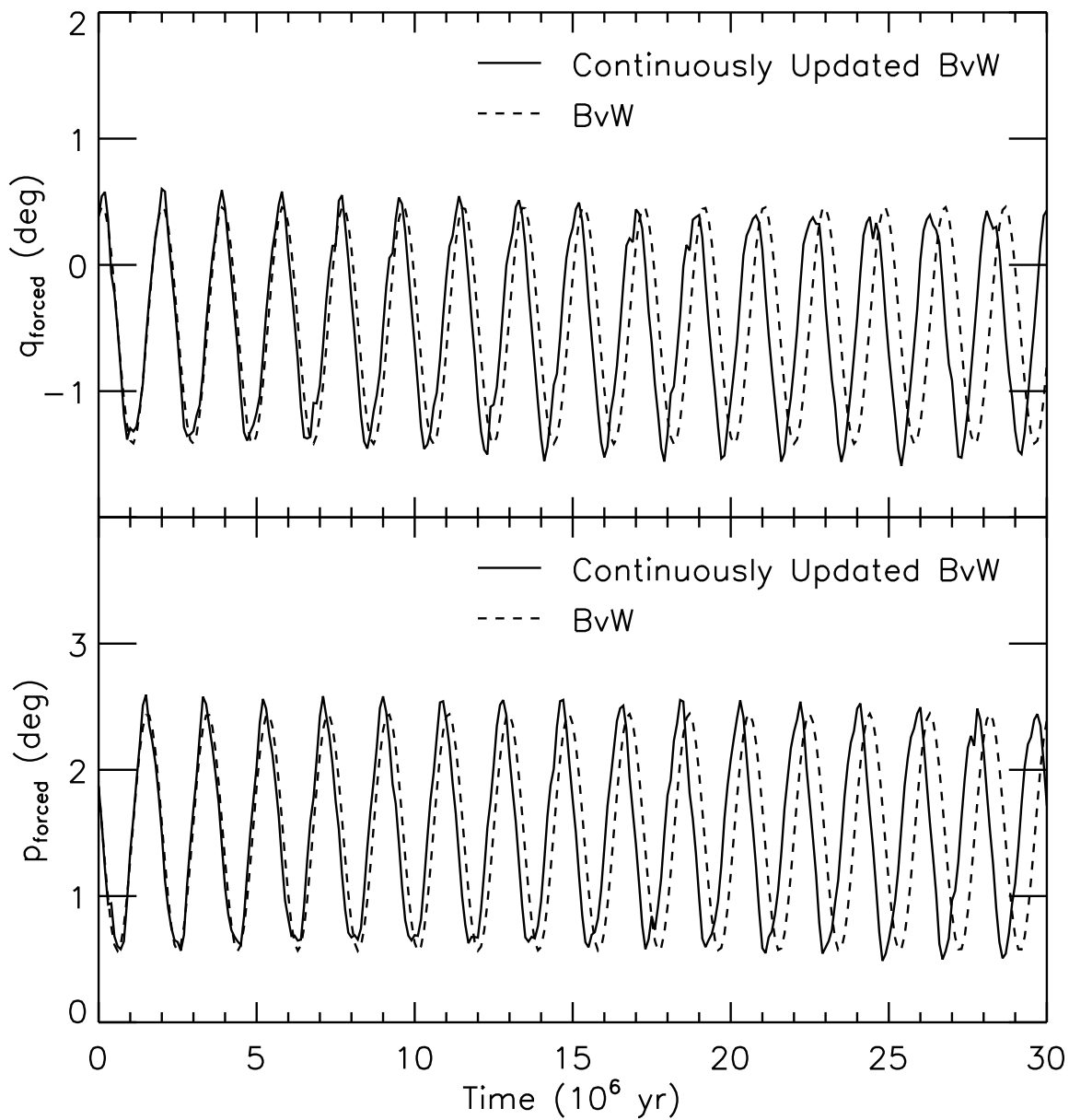}  
\caption{Components of the forced pole versus time, computed
using the BvW theory (dashed line)
and the continuously updated BvW theory (solid line).
The latter uses the results of numerical integrations at every timestep
to reset its parameters,
and is therefore essentially a numerical solution.
While the numerical solution eventually diverges from
the BvW solution, the two are qualitatively similar.
}  
\label{fig_ad}  
\end{figure}

These conclusions
are supported by our numerical integrations of giant planet and test
particle orbits lasting 4 Gyr.
These integrations show that a Kuiper belt object maintains a nearly
fixed orbital inclination with respect to the time-variable
Kuiper belt plane. This accords with linear theory, in which
the free inclination represents the undamped, constant amplitude
of a test particle's free oscillation. It may seem surprising
that the linear theory is vindicated in this regard while
it cannot accurately predict planetary motions and hence the future
location of the Kuiper belt plane. But the inaccuracies accrue only
slowly---Figure \ref{fig_ad} shows that it takes several precession
periods for the linear theory to diverge from numerical simulation
in predicting the location of the plane, and the analytic and numerical
solutions are always qualitatively similar. Referring back to the equations
of motion (\ref{ddotq})--(\ref{ddotp}), we see that if we take the planetary
forcing terms $q_j(t)$ and $p_j(t)$ to be given by the more realistic
numerical integration---instead of the inaccurate but qualitatively
similar analytic solution (\ref{q_j})--(\ref{p_j})---then the free
component of the motion (the homogeneous solution of the differential
equation) would still be given by (\ref{qfree})--(\ref{pfree}), irrespective
of the forced component (the inhomogenous solution).
Thus an object can keep its free inclination with respect to the forced
plane fixed, even though the location of the forced plane itself
cannot be forecast analytically.


\placefigure{fig_ten}

Currently the data on actual KBOs
are consistent with, but do not conclusively
verify, our theoretical finding that the
Kuiper belt plane warps by a few degrees to either side of the secularly
unstable gap.
Our analysis of the observations, like that of Elliot et al.~(2005),
suffers from large random errors. The study by Brown \& Pan (2004)
does not, but at the expense of including objects of all dynamical classes
and not following variations in the pole position with semi-major axis.
Quadrupling the 
sample size of low-$i$, low-$e$ objects at 38 AU, where there are currently
ten usable objects, can increase our confidence
in the reality of the warp to greater than 3$\sigma$.

Detecting ``Planet X'' via its influence on the Kuiper belt
plane will be substantially more challenging.
We calculate that a 100-$M_{\oplus}$ planet
with a semi-major axis of 300 AU and an orbital inclination with respect
to the ecliptic of 10 degrees would shift the forced poles in the 43--50 AU
region by 0.1 degree.

\acknowledgements
We thank Ruth Murray-Clay and Mike Brown for discussions,
and Chris Culter for suggesting this undergraduate research collaboration.
An anonymous referee provided numerous helpful comments.
This work was supported by NSF grant AST-0507805.

\end{document}